\documentclass[12pt,a4paper,dvips]{article}
\usepackage{epsfig}

\renewcommand{\topfraction}{0.99}
\renewcommand{\bottomfraction}{0.99}
\renewcommand{\textfraction}{0.01}
\begin{document}

\title{\bf Calculation of fluxes of charged particles and neutrinos from atmospheric showers}

\author{\bf{V.Plyaskin\thanks{{\it E-mail address:} v.plyaskine@cern.ch}} \\
\it{Institute of Theoretical and Experimental Physics,}\\
\it{Moscow,Russia }\\}

\date{March 18, 2003.}

\maketitle

\begin{abstract}
The results on the fluxes of charged particles and neutrinos from
a 3-dimensional (3D) simulation of atmospheric showers are presented.
An agreement of calculated fluxes with data on charged particles
from the AMS and CAPRICE detectors is demonstrated.
Predictions on neutrino fluxes at different experimental sites  
are compared with results from other calculations. 

\end{abstract}

\section{\bf Introduction}
%\indent 
The interpretation of the data on atmospheric neutrinos from
different experiments \cite{bib-SKam,bib-Skam1,bib-MACRO} relies on calculations of neutrino fluxes.
The many calculations made in recent years 
\cite{bib-NFC,bib-Barr,bib-Bug,bib-Hon,bib-Lipa,bib-Naum,bib-PLY,bib-BattF,bib-Wentz1} have been done using 
approaches invoking different models of the Earth's magnetic field, different 
primary spectra of cosmic rays and methods of tracing them in the
Earth's magnetic field, as well as various models of production of secondaries
in the inelastic interactions and the way the Earth's magnetic field is
influencing (or not) the propagation of these secondaries in the atmosphere.
(For a review see \cite{bib-GaHo}).  

In view of fundamental importance of results of calculations for the interpretations
of the atmospheric neutrino anomaly, it is acknowledged that a comprehensive simulation 
of atmospheric showers is needed. However, because of big computing power required to
perform this task, most calculations resort to some, presumably well argumented 
simplifications. In the previous work \cite{bib-PLY}\footnote[1]{A technical error in
normalization of neutrino fluxes in Phys.Lett.B 516 is corrected for in hep-ph/0103286 v3.} 
an attempt has been made to
avoid these simplifications. The calculations were based on the recent data from
the AMS experiment \cite{bib-AMS} and demonstrated a good agreement with experimental
results in terms of fluxes of charged secondaries produced in atmospheric showers
\cite{bib-AMS,bib-mu}. 

The present study follows the same approach with some changes.
Firstly, the rigidity range of generated primary cosmic rays is extended from 100 to 500 GV.
Secondly, the contribution from helium nuclei is simulated directly in He interactions with
atmosphere and not using the results from proton interactions. Finally, the statistics of 
this study is about 3.5 times bigger and corresponds to $\sim$ 2.2 ps exposure.

\section{\bf Description of the model}
The calculation is based on the GEANT3/GHEISHA package \cite{bib-GEANT}
adapted to the scale of cosmic ray travel and interactions.
The Earth is modeled to be a sphere of 6378.14 km radius of a uniform density of 
5.22 g/$\rm cm^3$. The atomic and nuclear properties
of the Earth are taken to be those of Ge - the closest in density (5.32 g/$\rm cm^3$) element.
The Earth's atmosphere is modeled by 1 km thick variable density layers of air extending up
to 71 km from the Earth's surface. The density change with altitude is calculated using Standard
Atmosphere Calculator \cite{bib-atm}. The Earth's magnetic field for the year 1998 is 
calculated according to World Magnetic Field Model (WMM-2000) \cite{bib-WMM} with 6 degrees of 
spherical harmonics. 
 
The flux of primary protons and helium nuclei in the Solar system is parametrized on the basis
of the AMS data \cite{bib-AMS}. At the rigidities below several GV the spectrum 
is corrected for the solar activity according to the 11 year solar cycle \cite{bib-p6}.
Isotropically emitted from a sphere of 10 Earth's radii, the primary cosmic particles with 
kinematic parameters compatible with those reaching the Earth are traced in the magnetic
field until they interact with the atmosphere. The production of secondaries resulting from 
interactions of He nuclei is treated using the superposition approximation \cite{bib-Schu}.
The leading primary and the secondary particles produced in the interactions are traced 
in the magnetic field until they go below 0.125 GeV/c.  

\section{\bf Comparison with experimental data}

\begin{figure}[hp]
\begin{center}\mbox{
\epsfig{file=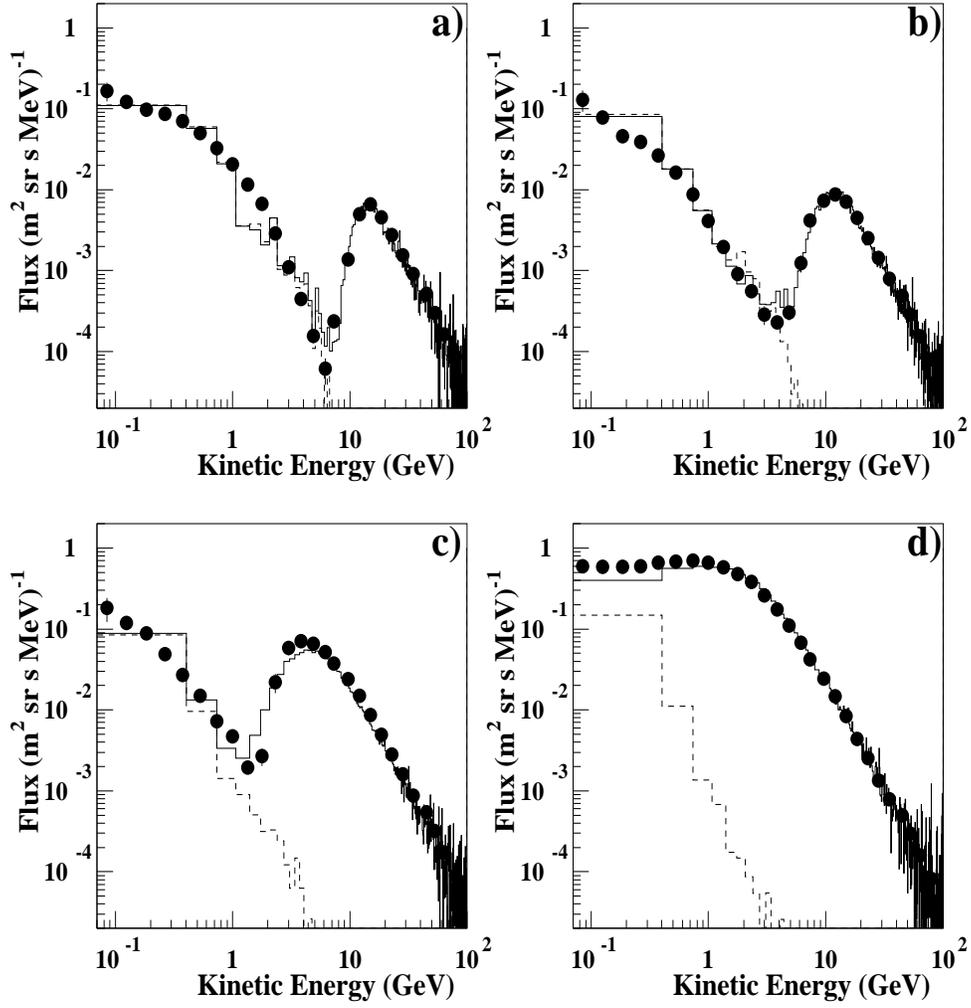,height=15 cm,width=15cm}}
\caption{%
Proton flux in AMS at different magnetic latitudes ($|\Theta_M|$). Solid line - flux from the top, 
dashed line - flux from the bottom.
a) $|\Theta_M|<$0.2 rad, b) 0.2$<|\Theta_M|<$0.5 rad, c) 0.5$<|\Theta_M|<$0.8 rad,
c) 0.8$<|\Theta_M|<$1.1 rad. The dots are the AMS measurements \cite{bib-AMS}.}
\label{fig:508m}
\end{center}
\end{figure}

All results in this study are obtained directly without any renormalization.
To compare the calculated fluxes of charged particles with the AMS measurements \cite{bib-AMS} 
the overall fluxes produced in the simulation are restricted to the AMS acceptance 
during the zenith facing flight period.
Fig.\ref{fig:508m} shows the spectra of protons for different positions of AMS with respect
to the magnetic equator. As in the previous study the result of simulation correctly
reproduces the spectra of both primary and secondary protons. 

\newpage

\newpage

\begin{figure}[hp]
\begin{center}\mbox{
\epsfig{file=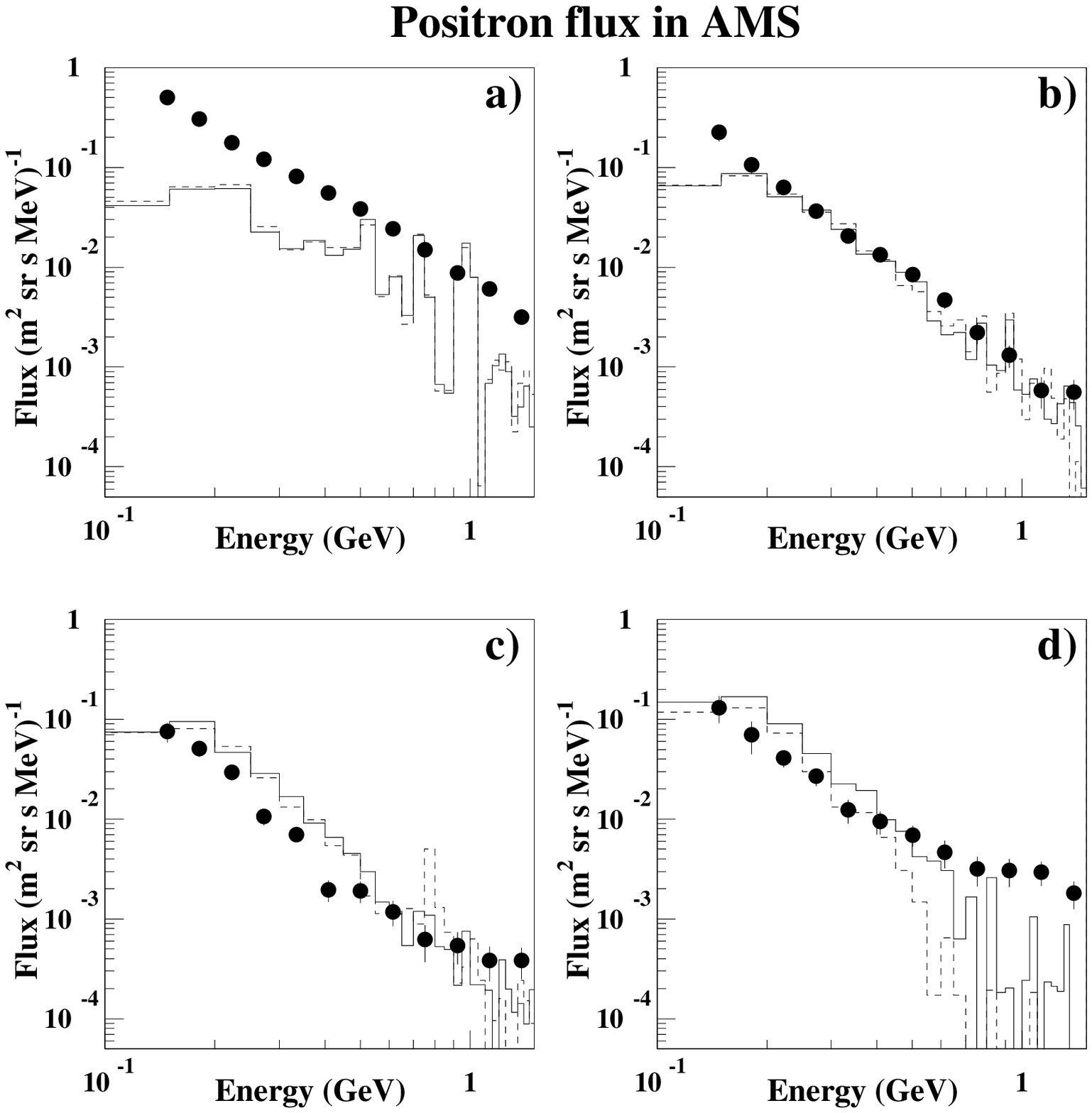,height=15 cm,width=15cm}}
\caption{%
Secondary positron flux in AMS at different magnetic latitudes ($|\Theta_M|$). Solid line - flux from the top, 
dashed line - flux from the bottom.
a) $|\Theta_M|<$0.3 rad, b) 0.3$<|\Theta_M|<$0.6 rad, c) 0.6$<|\Theta_M|<$0.8 rad,
c) 0.8$<|\Theta_M|<$1.1 rad. The dots are the AMS measurements \cite{bib-AMS}. The enhancement at higher energies of the measured flux
at high magnetic latitude is due to primary cosmic positrons detected in the lower magnetic cutoff
region.}
\label{fig:510m}
\end{center}
\end{figure}

\newpage
\begin{figure}[hp]
\begin{center}\mbox{
\epsfig{file=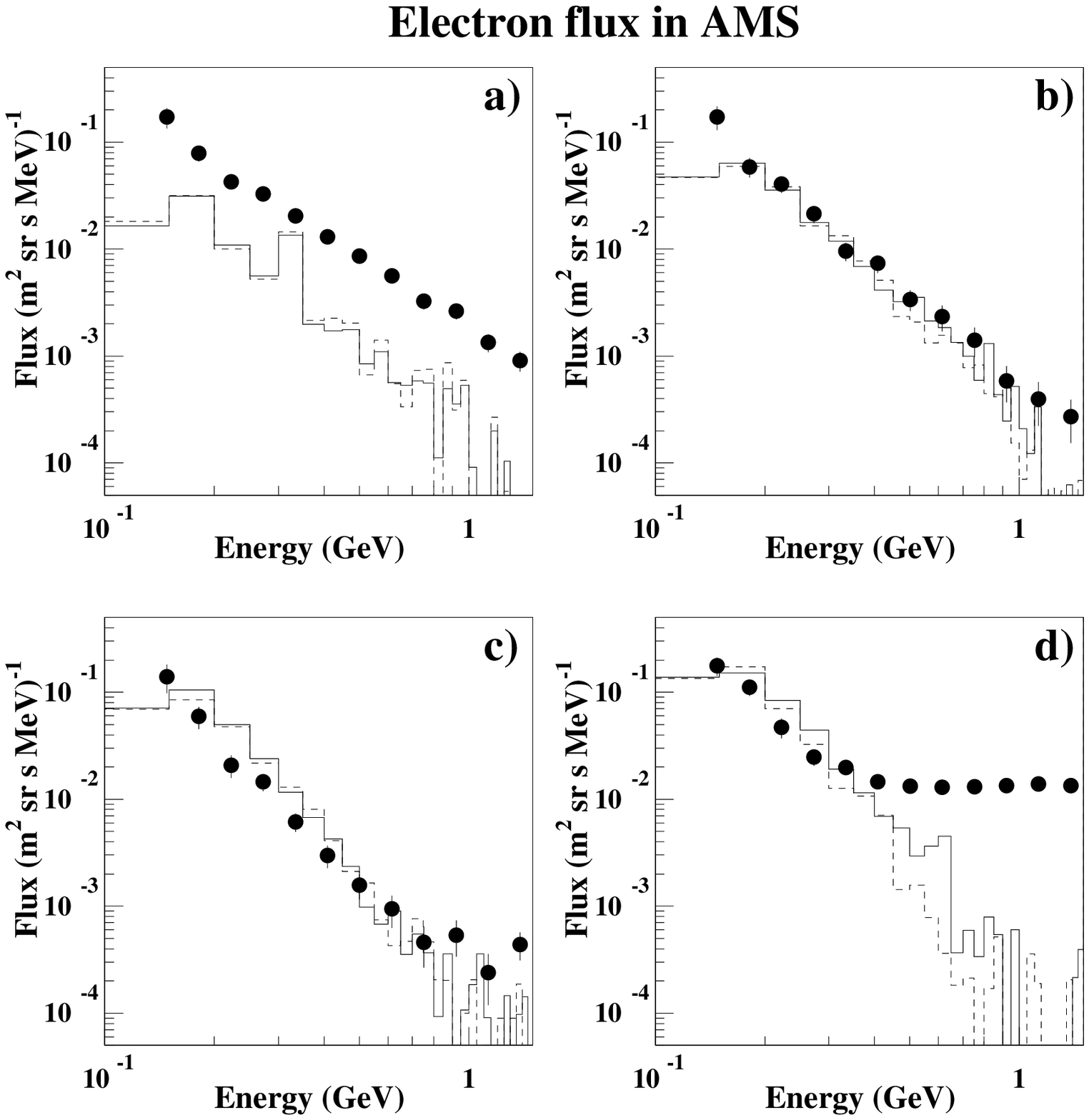,height=15 cm,width=15cm}}
\caption{%
Secondary electron flux in AMS at different magnetic latitudes ($|\Theta_M|$). Solid line - flux from the top, 
dashed line - flux from the bottom.
a) $|\Theta_M|<$0.3 rad, b) 0.3$<|\Theta_M|<$0.6 rad, c) 0.6$<|\Theta_M|<$0.8 rad,
d) 0.8$<|\Theta_M|<$1.1 rad. The dots are the AMS measurements \cite{bib-AMS}. The enhancement at higher energies of the measured flux
at high magnetic latitude is due to primary cosmic electrons detected in the lower magnetic cutoff
region.}
\label{fig:511m}
\end{center}
\end{figure}

\newpage
\begin{figure}[ht]
\begin{center}
\mbox{\epsfig{file=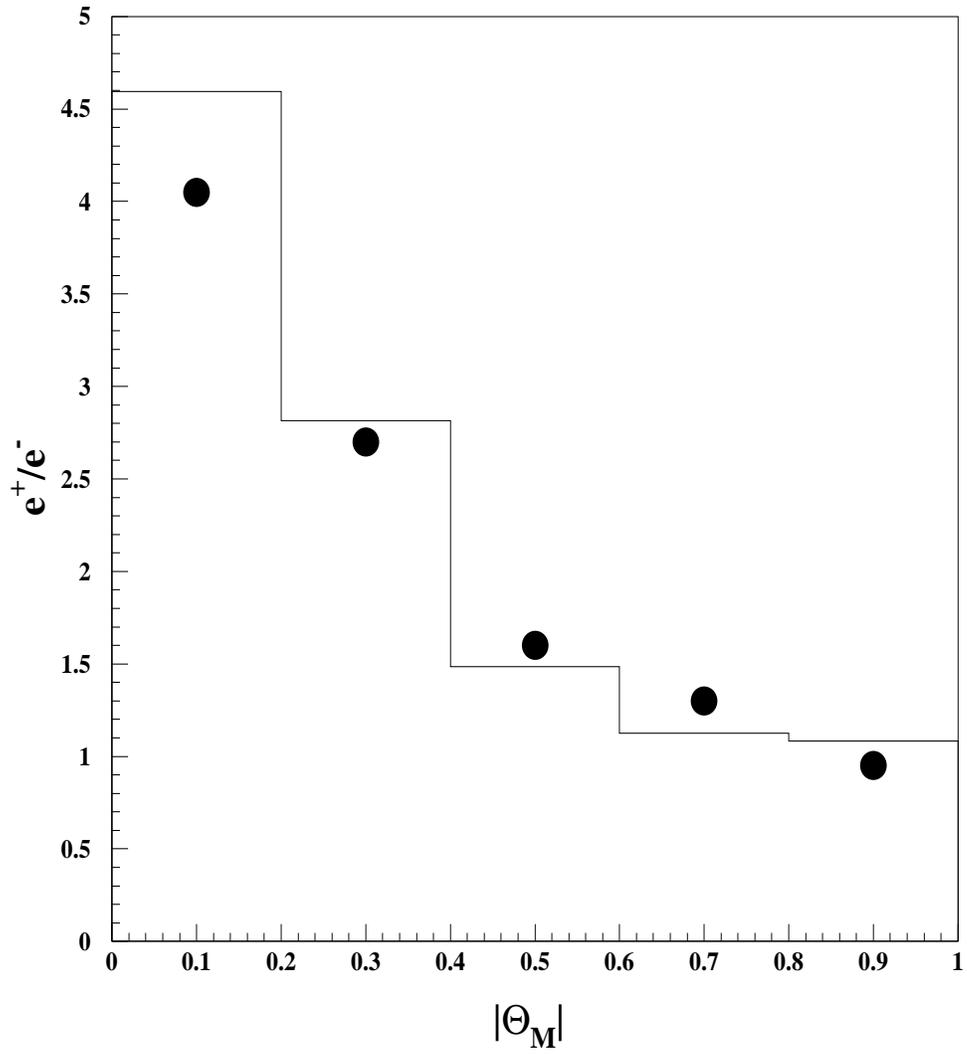,height=15.0 cm,width=15.0 cm}}
\caption{%
Calculated dependence on magnetic latitude of positron/electron flux ratio.
The dots are the AMS measurements \cite{bib-AMS}.}
\label{fig:512m}
\end{center}
\end{figure}

\newpage

\begin{figure}[hp]
\begin{center}\mbox{
\epsfig{file=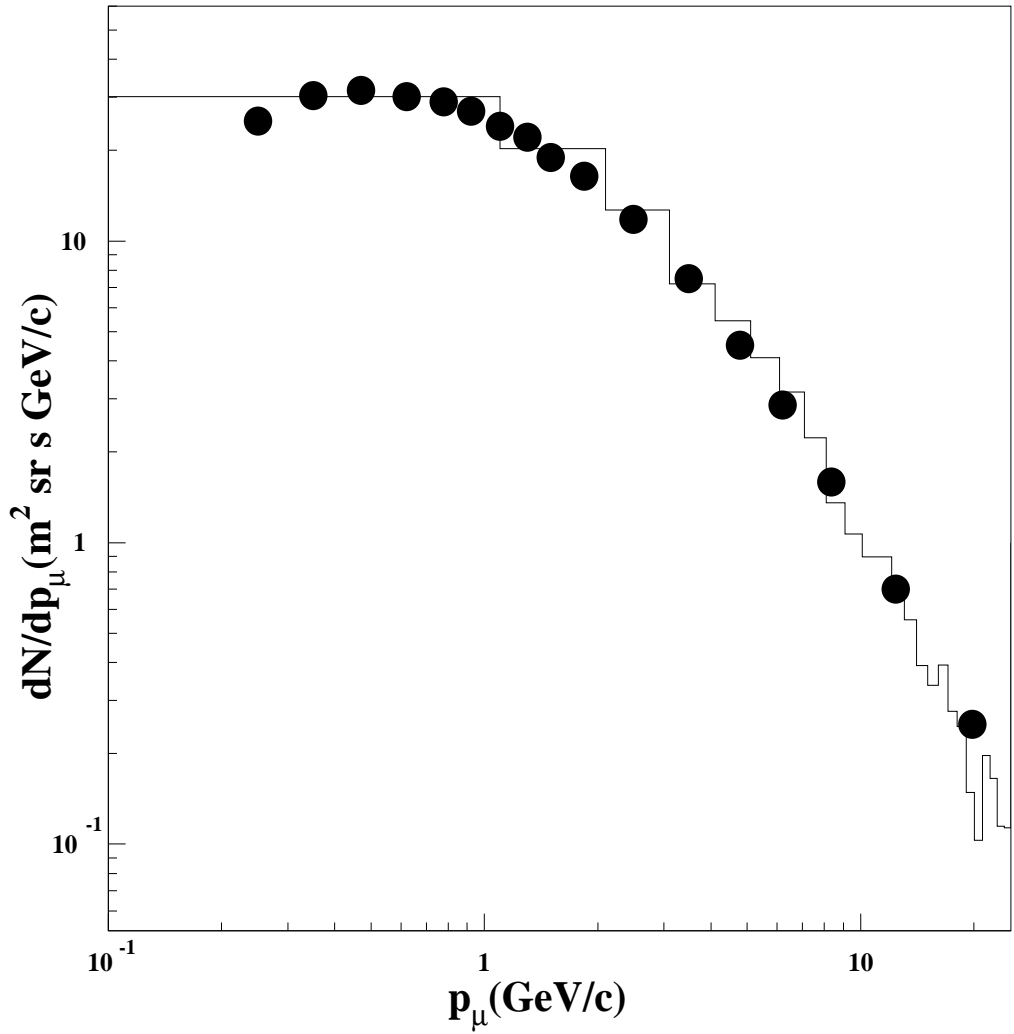,height=15 cm,width=16cm}}
\caption{%
Vertical muon flux at the sea level. Histogram: this study.
Circles: data from CAPRICE 94 \cite{bib-CAPRICE}.} 
\label{fig:muons}
\end{center}
\end{figure}

\newpage
The simulated positron and electron spectra shown in Figs.\ref{fig:510m} and \ref{fig:511m} 
are in good agreement with those obtained in the AMS experiment \cite{bib-AMS} allowing for
still insufficient statistics of spiraling positrons and electrons in the equator region, where
the spiraling secondaries dominate. For detailed discussion of the behaviour of the secondaries 
see \cite{bib-PLY}. 
The measured relative rate of secondary leptons
and their dependence on the magnetic latitude is well reproduced by the simulation 
(Fig.\ref{fig:512m}). 

The histogram in Fig.\ref{fig:muons} shows the flux of cosmic muons within the acceptance
of the CAPRICE 94 experiment. The calculated flux is in good agreement with the measured one 
reported in \cite{bib-CAPRICE}.

A good agreement between the predictions of fluxes of charged secondaries provided by this 
calculation and the measurements is a strong indication that the fluxes of neutrino,
which are produced in the same decays as the charged particles (see Fig.\ref{fig:nuprod}) 
are simulated correctly.

\section{\bf Atmospheric neutrino fluxes}

\begin{figure}[hb]
\begin{center}\mbox{
\epsfig{file=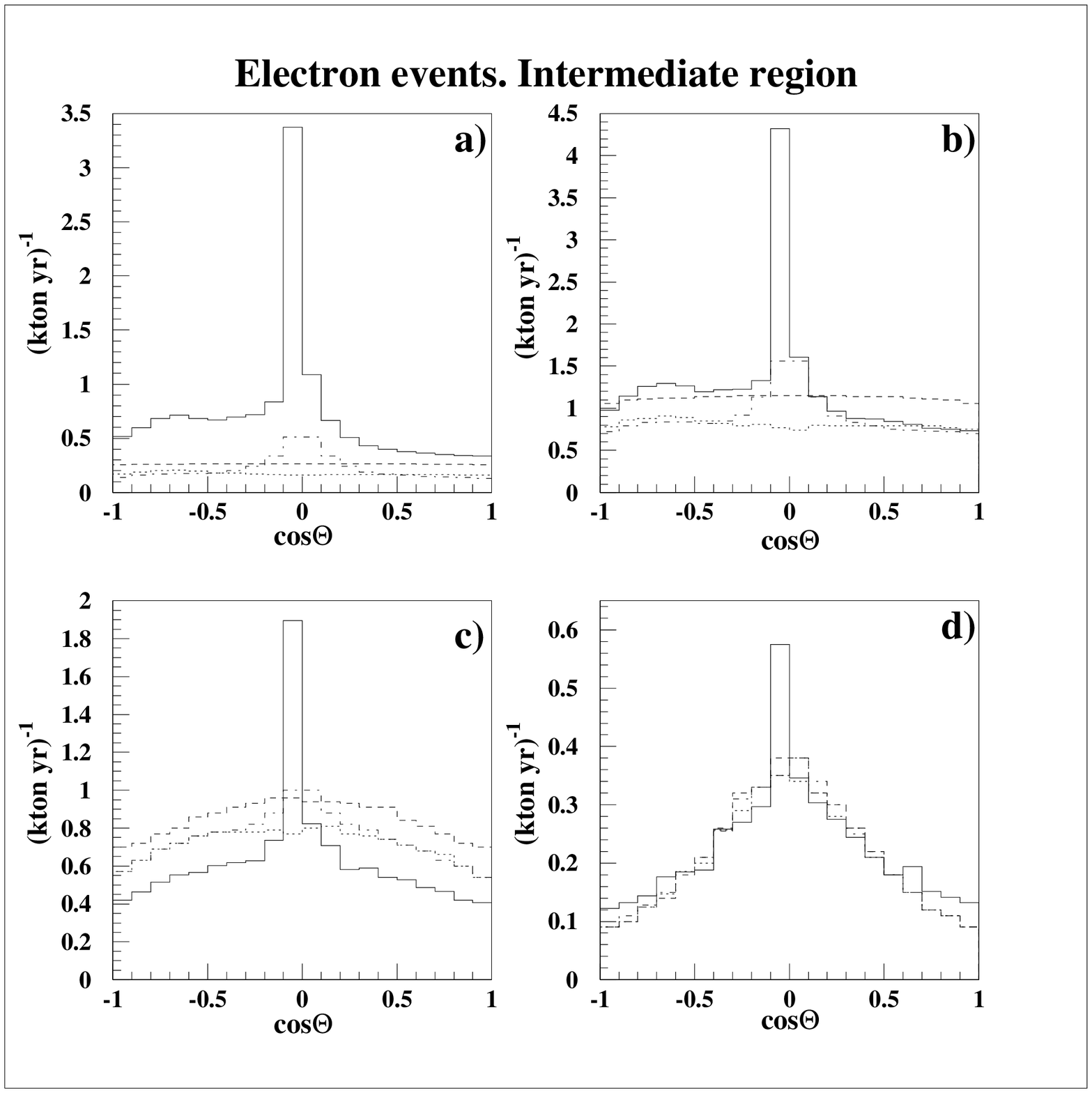,height=15 cm,clip,width=16.3cm,clip}}
\caption{%
Average nadir angle distribution for e-like events for detectors located in positions on the Earth  
with magnetic latitude sin($\rm \Theta_M$)= [0.2,0.6]. a) $\rm E_{\nu}$=0.1-0.31 GeV, 
b) $\rm E_{\nu}$=0.31-1.0 GeV, c) $\rm E_{\nu}$=1.0-3.1 GeV, d) $\rm E_{\nu}$=3.1-10.0 GeV.
From Ref.\cite{bib-Lipa} : dashed line - 1D calculation without geomagnetic effects; 
dotted line - 1D calculation; dash-dotted line - 3D calculation. Solid line -this work.}
\label{fig:elint}
\end{center}
\end{figure}

\begin{figure}[hb]
\begin{center}\mbox{
\epsfig{file=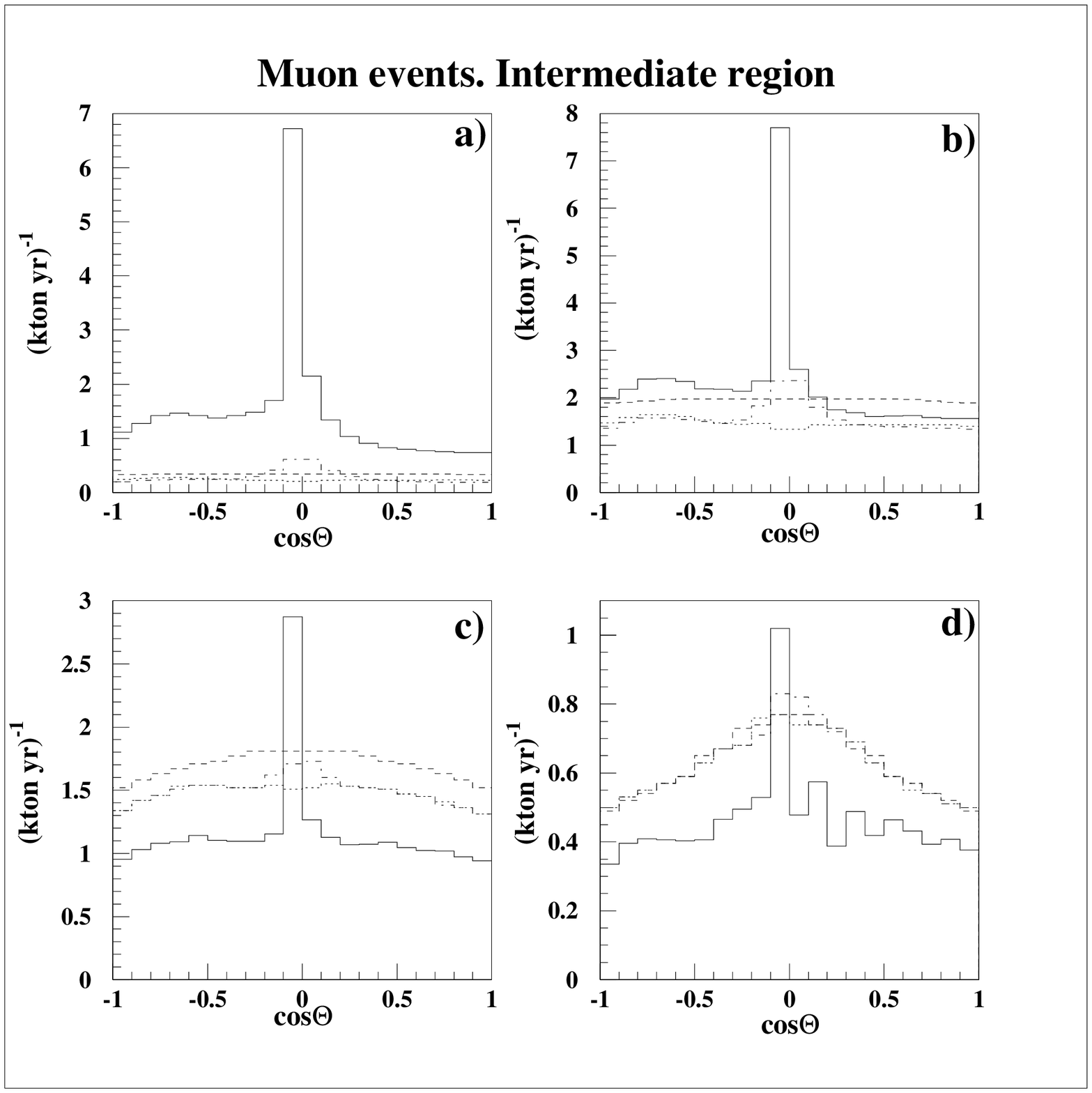,height=15 cm,clip,width=16.3cm,clip}}
\caption{%
Average nadir angle distribution for $\mu$-like events for detectors located in positions on the Earth  
with magnetic latitude sin($\rm \Theta_M$)= [0.2,0.6]. a) $\rm E_{\nu}$=0.1-0.31 GeV, 
b) $\rm E_{\nu}$=0.31-1.0 GeV, c) $\rm E_{\nu}$=1.0-3.1 GeV, d) $\rm E_{\nu}$=3.1-10.0 GeV.
From Ref.\cite{bib-Lipa} : dashed line - 1D calculation without geomagnetic effects; 
dotted line - 1D calculation; dash-dotted line - 3D calculation. Solid line -this work.}
\label{fig:muint}
\end{center}
\end{figure}

\begin{figure}[hb]
\begin{center}\mbox{
\epsfig{file=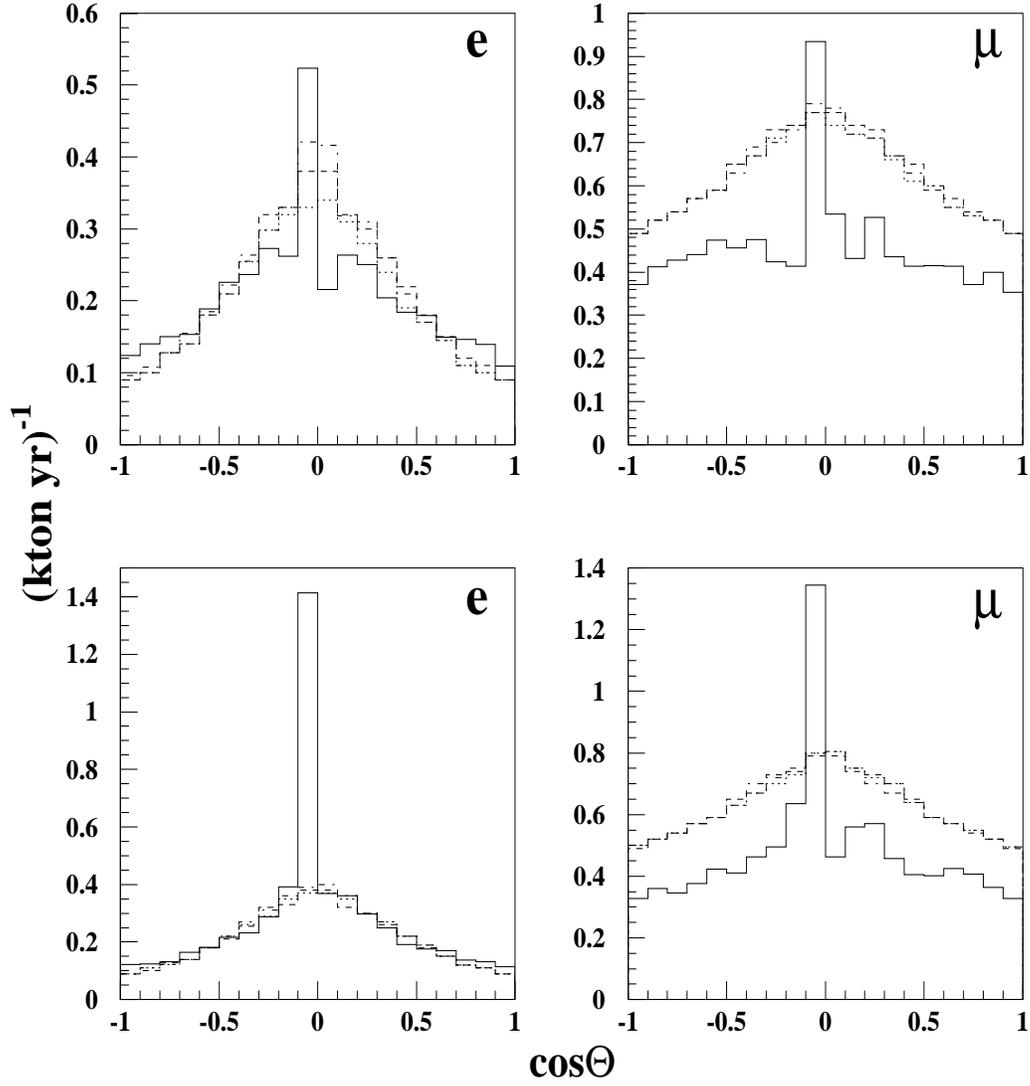,height=16 cm,width=16 cm}}
\caption{%
Average nadir angle distribution for e-like and $\mu$-like events 
induced by neutrinos with energy $\rm E_{\nu}$=3.1-10.0 GeV.
For detectors located in positions on the Earth with magnetic latitude:
top - equator region, sin($\rm \Theta_M$)= [-0.2,0.2]; bottom - polar region, sin($\rm \Theta_M$)= [0.6,1.0].
From Ref.\cite{bib-Lipa} : dashed line - 1D calculation without geomagnetic effects; 
dotted line - 1D calculation; dash-dotted line - 3D calculation. Solid line -this work.}
\label{fig:lips}
\end{center}
\end{figure}

Figs.\ref{fig:elint},\ref{fig:muint},\ref{fig:lips} 
show the angular distributions of neutrino event rate in different regions according to 
the geomagnetic latitude calculated using the neutrino fluxes obtained in this work.
These distributions are the result of convolution of neutrino
fluxes with the neutrino cross sections \cite{bib-LipConv},\cite{bib-Conv1},\cite{bib-Conv2}
and do not include detector related effects. The actual amount of detected neutrinos from 
those produced just under the Earth's surface, which is modeled as a simple sphere, depends 
on the actual detector position
(depth underground, surrounding landscape). 
Compared to \cite{bib-Lipa} the event rates from this work, given in 
Figs.\ref{fig:elint},\ref{fig:muint} for the magnetic latitude region with
sin($\rm \Theta_M$)= [0.2,0.6], are systematically higher for very low, 
$\rm E_{\nu}$=0.1-0.31 GeV, energy neutrinos, and are similar in the energy
range of $\rm E_{\nu}$=0.31-1.0 GeV. For higher energies of $\rm E_{\nu}$=1.0-3.1 GeV this
work predicts event rates which are at the level of 0.7-0.8 of the 3D calculation result 
of \cite{bib-Lipa}. The important difference with previous calculations is clearly visible 
in the highest neutrino energy range of $\rm E_{\nu}$=3.1-10.0 GeV. On the one hand the 
angular distributions and the absolute event rates for e-like events are practically the 
same for all types of calculations. On the other
hand the integral event rate for $\rm \mu$-like events is only about 0.6 of those predicted 
in \cite{bib-Lipa} and the angular distribution of this events, apart form the angular region
close to the horizontal direction, is much flatter. This tendency is typical for 
high energy neutrino events at all
magnetic latitude regions (see Fig.\ref{fig:lips}).  

\begin{figure}[hp]
\begin{center}\mbox{
\epsfig{file=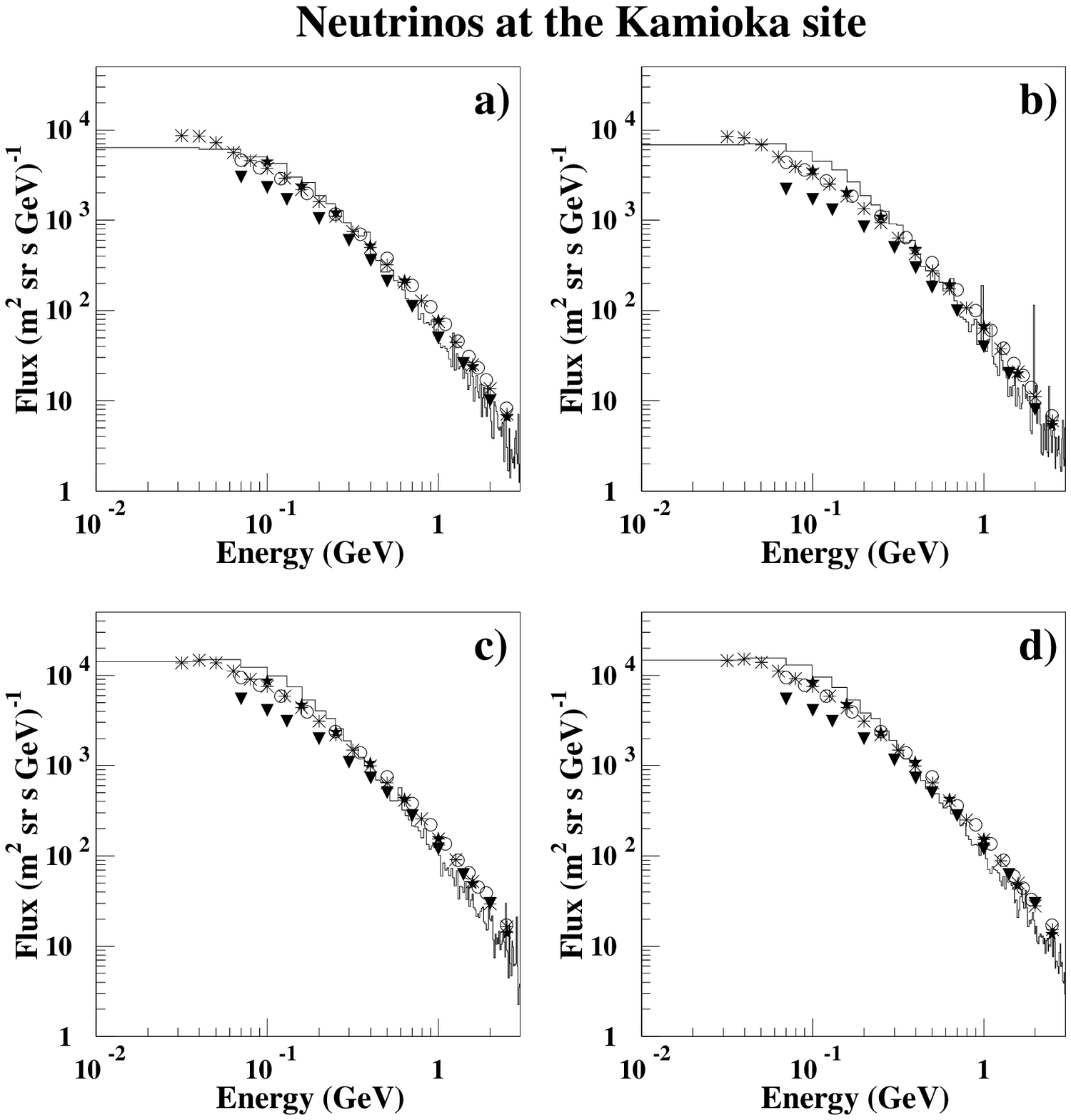,height=15 cm,width=16.3cm}}
\caption{%
Differential energy spectra of atmospheric neutrino (averaged over all directions) 
at the Kamioka site. 
a) $\nu_{e}$, b) $\bar{\nu}_{e}$, c) $\nu_{\mu}$, d) $\bar{\nu}_{\mu}$. 
Asterisks from Ref. \cite{bib-Hon},
open circles from Ref. \cite{bib-Barr}, triangles from Ref. \cite{bib-Bug}, 
stars from Ref.\cite{bib-Naum}. Histograms - this work.}
\label{fig:515a}
\end{center}
\end{figure}

\begin{figure}[hp]
\begin{center}\mbox{
\epsfig{file=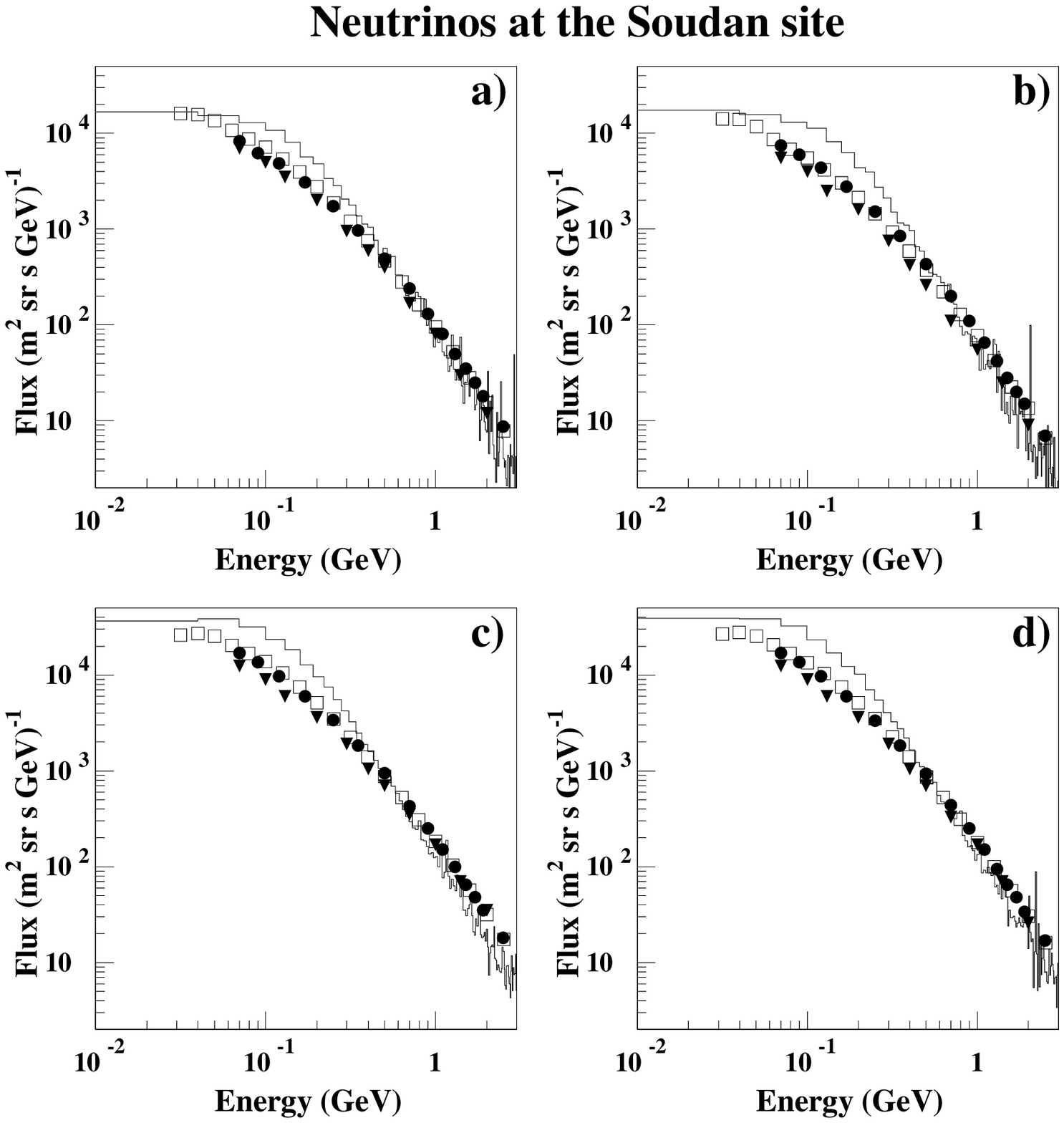,height=15 cm,width=16.3cm}}
\caption{%
Differential energy spectra of atmospheric neutrino (averaged over all directions)
 at the Soudan site. 
a) $\nu_{e}$, b) $\bar{\nu}_{e}$, c) $\nu_{\mu}$, d) $\bar{\nu}_{\mu}$.
Squares from Ref.\cite{bib-Hon},
dots from Ref.\cite{bib-Barr}, triangles from Ref.\cite{bib-Bug}. Histograms - this work.}
\label{fig:517a}
\end{center}
\end{figure}

\begin{figure}[hp]
\begin{center}\mbox{
\epsfig{file=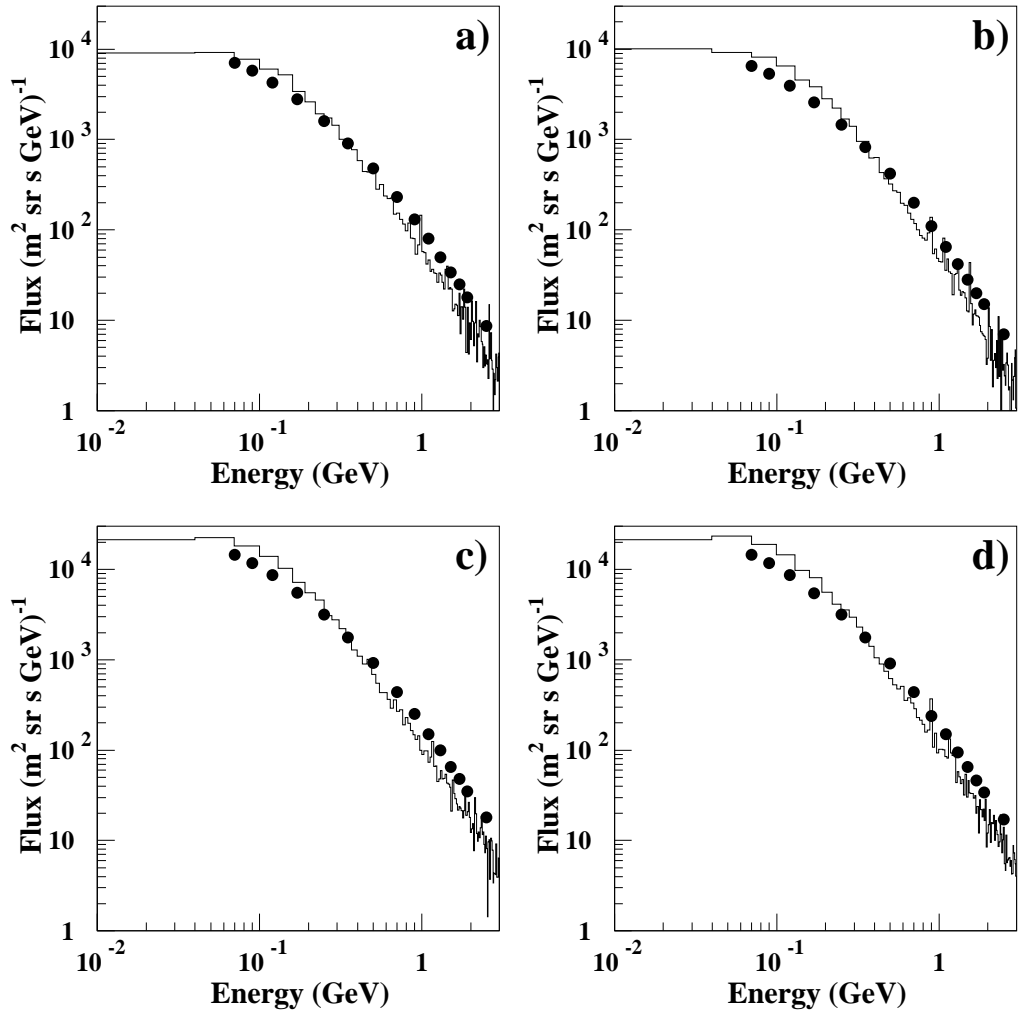,height=15 cm,width=16.3cm}}
\caption{%
Differential energy spectra of atmospheric neutrino (averaged over all directions)
at the Gran Sasso site. 
a) $\nu_{e}$, b) $\bar{\nu}_{e}$ c) $\nu_{\mu}$, d) $\bar{\nu}_{\mu}$. 
Dots from Ref.\cite{bib-Barr}. Histograms - this work.}
\label{fig:519a}
\end{center}
\end{figure}

The neutrino flux at different experimental sites is calculated for the area $\pm12^o$ in latitude and
longitude with the center at ($35^o~N,~142^o E$);  ($48^o~ N,~98^o~ W$) and ($42^o~ N,~13^o E$) and the
depth underground of 520 (2700 mwe), 700 and 750 m for Kamioka, Soudan and Gran Sasso sites 
correspondingly.
Fig.\ref{fig:515a} shows the averaged over all directions differential energy spectra of 
different neutrinos and antineutrinos at the Kamioka site. 
The spectra for Soudan and Gran Sasso sites are presented in Figs.\ref{fig:517a} and \ref{fig:519a} 
correspondingly. 

\newpage

\newpage

\newpage
\newpage

\newpage

\newpage

\newpage

\newpage
\begin{figure}[hp] 
\begin{center}
\mbox{\psfig{file=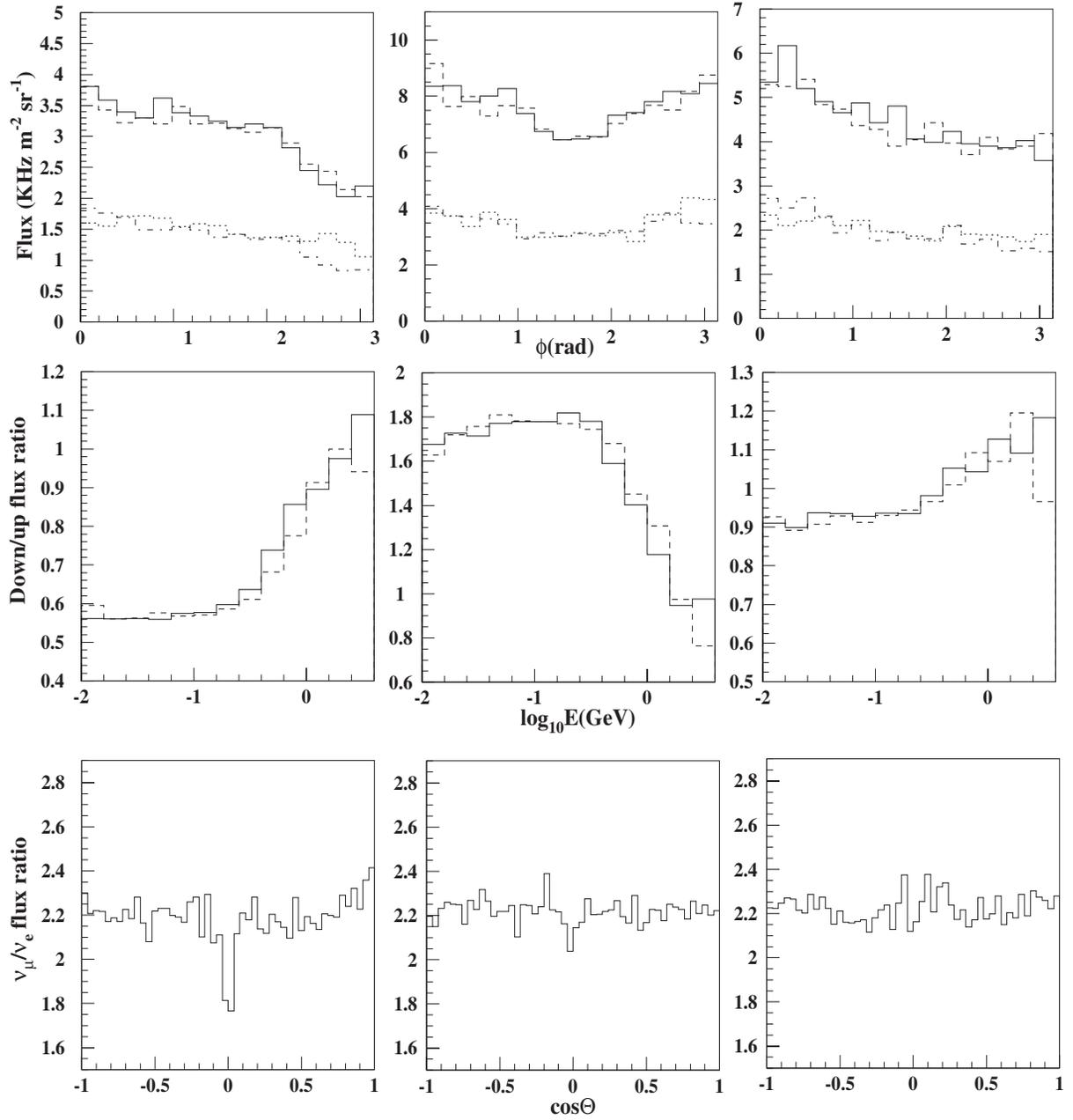,width=17cm,height=17cm}}
\end{center}
\caption{Neutrino flux at different sites. 
Left - Kamioka, center - Soudan, right -Gran Sasso.
Top: East/West distribution. Zero corresponds to the flux from the West.
$\nu_{\mu}$ -solid line, $\bar{\nu}_{\mu}$ - dashed lilne, $\nu_{e}$ -
dotted line, $\bar{\nu}_{e}$ - dash-dotted line.
Middle: Down/Up flux ratio. $\nu_{\mu}$+$\bar{\nu}_{\mu}$ -solid line,
$\nu_{e}$+ $\bar{\nu}_{e}$ -dashed line.
Bottom: $\nu_{\mu}$+$\bar{\nu}_{\mu}$/$\nu_{e}$+ $\bar{\nu}_{e}$ flux ratio.
} 
\label{fig:nuall}
\end{figure}

\newpage
\begin{figure}[hp] 
\begin{center}
\mbox{\psfig{file=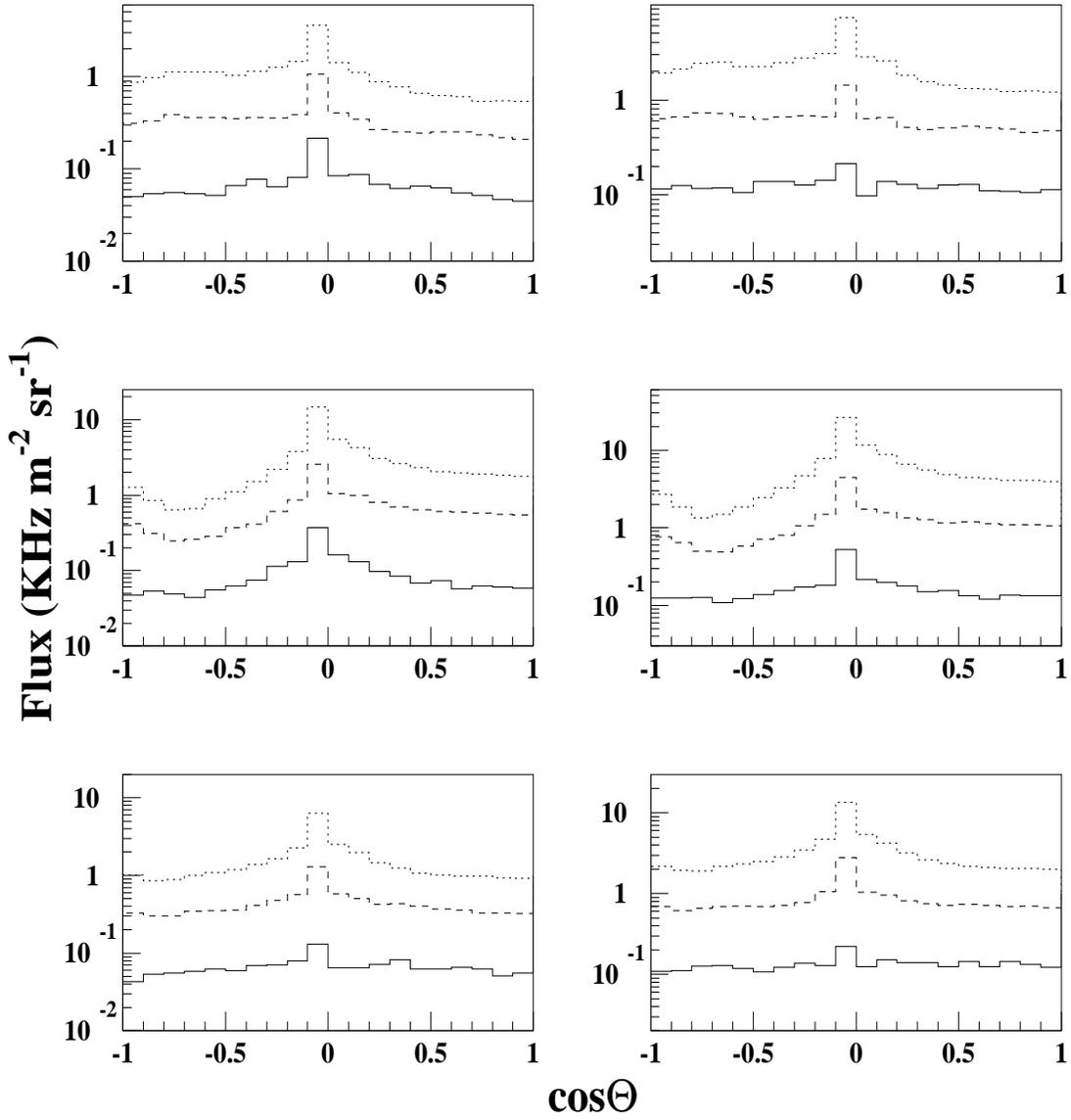,width=17cm,height=17cm}}
\end{center}
\caption{Angular dependence of neutrino flux at different sites. 
Top - Kamioka, middle - Soudan, bottom - Gran Sasso.
Left column - $\nu_{e}$+ $\bar{\nu}_{e}$, 
right column - $\nu_{\mu}$+$\bar{\nu}_{\mu}$.
Neutrino energy: Dotted line - 0.1$<E_{\nu}<$0.31 GeV; dashed line - 0.31$<E_{\nu}<$1.0 GeV;
solid line - $E_{\nu}>$1.0 GeV.
} 
\label{fig:nucos}
\end{figure}

\newpage
\begin{figure}[hp]
\begin{center}\mbox{
\epsfig{file=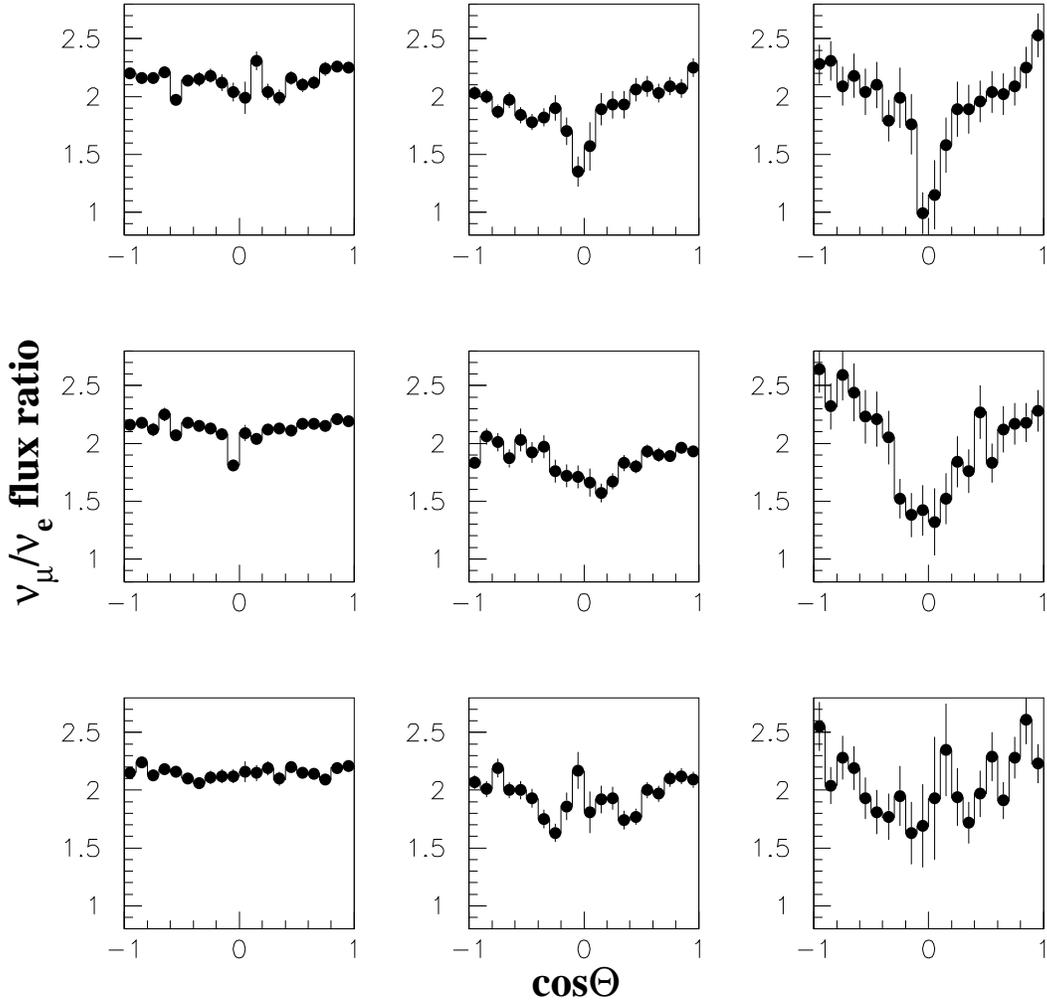,height=15 cm,width=16.3cm}}
\caption{%
Angular dependence of $\nu_{\mu}$+$\bar{\nu}_{\mu}$ / $\nu_{e}$+$\bar{\nu}_{e}$ 
flux ratio at different sites.
Neutrino energy: left column 0.1$<E_{\nu}<$0.31 GeV; central column 0.31$<E_{\nu}<$1.0 GeV;
right column $E_{\nu}>$1.0 GeV.
Top raw - Kamioka site; middle raw - Soudan site; bottom raw - Gran Sasso site.
The error bars give statistical errors. 
}
\label{fig:nurat}
\end{center}
\end{figure}

\newpage
In comparison with predictions of previous calculations made for this sites 
\cite{bib-Barr,bib-Bug,bib-Hon,bib-Naum} there is an excess of low energy and a deficit 
of high energy neutrinos. The same tendency is reported in \cite{bib-BattF}.  
The extent of these differences depends on the given site and is not the same for different kinds 
of neutrino.

From Fig.\ref{fig:nuall} one can see that there is a considerable difference both in East/West
and down/up asymmetry of the fluxes at different sites. At the same time the ratio of 
muon($\nu_{\mu}+\bar{\nu}_{\mu}$) to electron($\nu_{e}+\bar{\nu}_{e}$) neutrino fluxes,
dominated by low energy neutrinos, appears to be constant at all sites and 
independent on the direction of neutrinos with respect to the surface of the Earth.
The situation is much different if one considers different neutrino energy intervals 
(Fig.\ref{fig:nucos}). As it 
is shown in Fig.\ref{fig:nurat} the muon/electron neutrino flux ratio is constant only for
low energy neutrinos. With the increase of neutrino energy the ratio behaves differently
at different sites, although there is a clear tendency towards smaller values of muon/electron neutrino
flux ratio for directions approaching horizontal. This effect, reported in our previous work \cite{bib-PLY}
is confirmed by the present higher statistics calculations. 

\section{\bf Discussion}
The neutrino fluxes predicted in this work for different locations have the energy spectra, 
relative intensity and angular distributions different from those obtained by other authors.
There are several points in calculations which can explain these differences.
\subsection{\bf Primary cosmic ray flux}      
\noindent
It is recognized, that the primary cosmic ray spectrum is a major source of uncertainty
in calculations of atmospheric neutrino fluxes. 
As it was demonstrated in section 3, the primary proton flux calculated using the forward
tracing method fits well the AMS data. In this section we present the results on primary 
proton flux at different magnetic latitudes obtained using the backward tracing method 
commonly used as a standard procedure in atmospheric neutrino flux calculations.
In this study the protons with the interstellar rigidity spectra are 
isotropically emitted from a sphere situated at the the top of the atmosphere at 70 km from
the Earth's surface and are traced following the backward path of a particle with the same 
rigidity but of the opposite charge. Only particles back-traced to a large distance 
(10 Earth's radii) are considered as following an "allowed" trajectory.

\begin{figure}[hp] 
\begin{center}
\mbox{\psfig{file=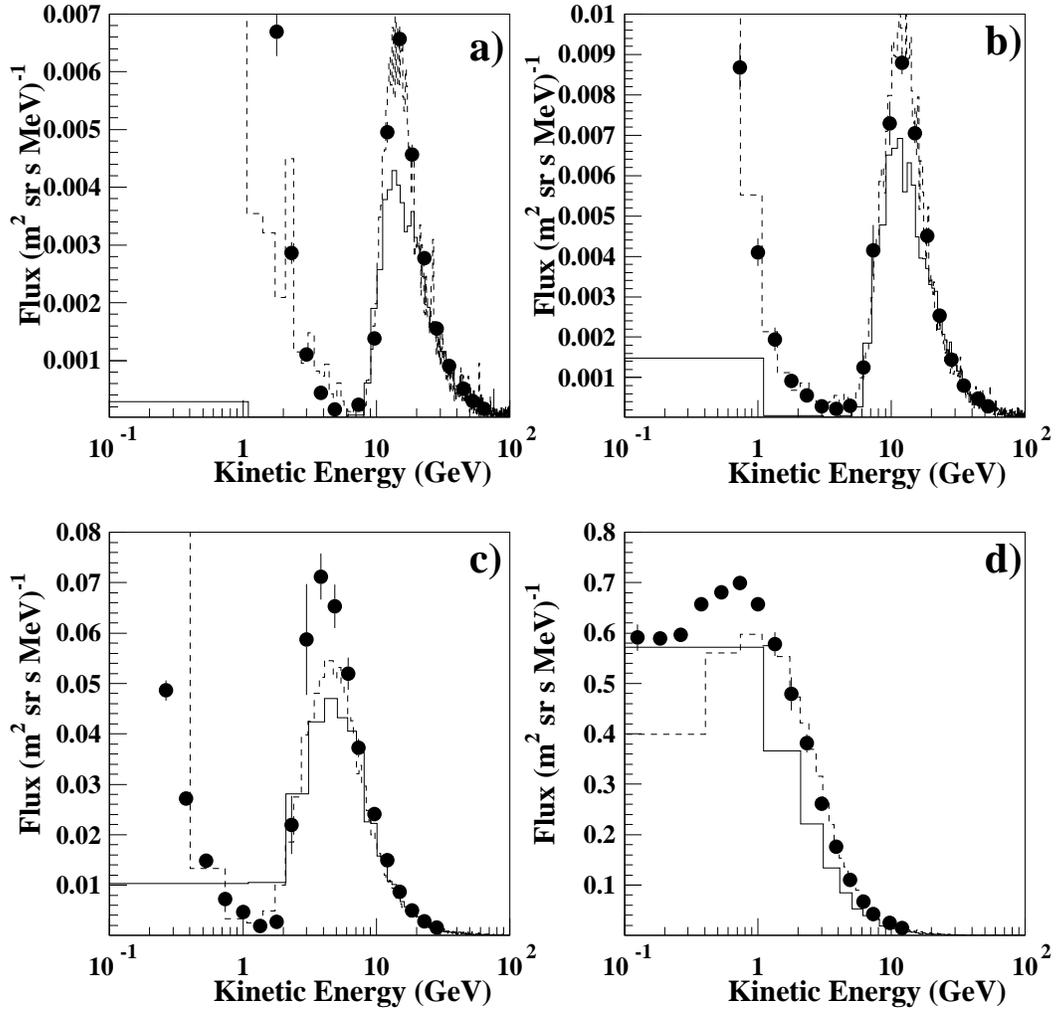,width=15cm,height=15cm}}
\end{center}
\caption{%
Proton flux in AMS at different magnetic latitudes ($|\Theta_M|$). Solid line - back tracing, 
dashed line - forward tracing.
a) $|\Theta_M|<$0.2 rad, b) 0.2$<|\Theta_M|<$0.5 rad, c) 0.5$<|\Theta_M|<$0.8 rad,
c) 0.8$<|\Theta_M|<$1.1 rad. The dots are the AMS measurements \cite{bib-AMS}.}
\label{fig:back4}
\end{figure}

The flux of primary protons with "allowed" trajectories passing through a detection plane 
from the top hemisphere is shown in Fig.\ref{fig:back4} for different magnetic latitude 
regions. The detection sphere is assumed to be at 400 km, corresponding to the AMS flight 
altitude.

\newpage

\newpage

A comparison with AMS measurements (see Fig.\ref{fig:back4}) 
shows that the cutoff value at different magnetic latitudes is predicted correctly
for both backward and forward tracing.
However, at variance with predictions obtained with forward tracing, the flux 
of "allowed" backtraced protons  approaches the measured values only 
above 40 GeV and is far below elsewhere \cite{bib-PLY1}. 
For protons responsible for production of low energy neutrinos it is about
2 times less then in the experiment. This deficit in primary flux intensity could be a reason 
for the lower fluxes and, consequently event rates, predicted for low energy neutrinos in 
previous calculations as compared with the present one.

\begin{figure}[htb]
\begin{center}\mbox{
\epsfig{file=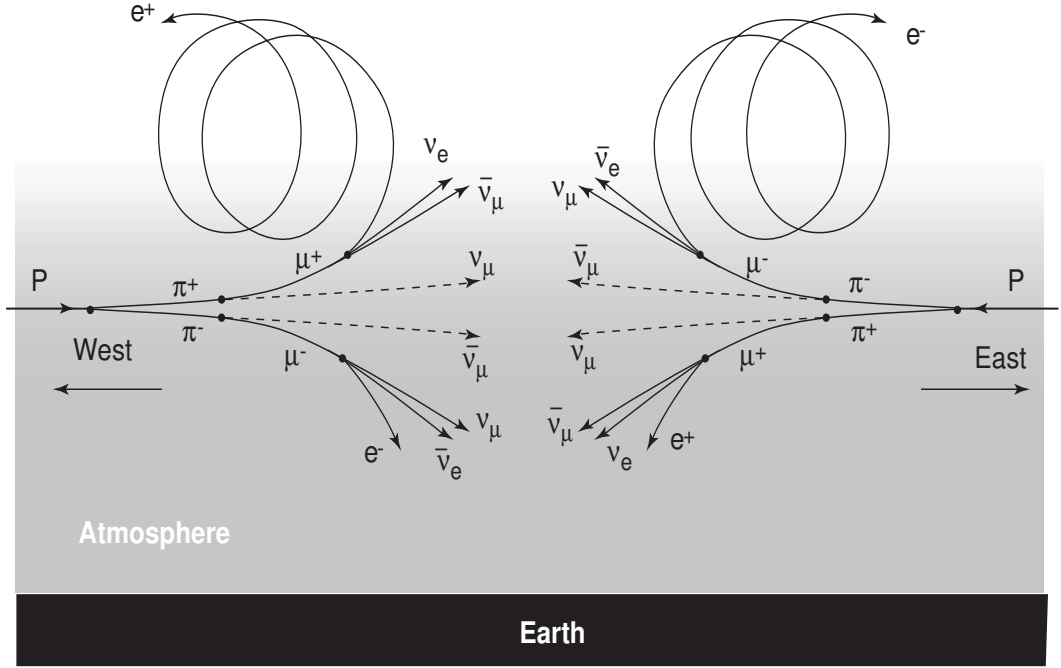,height=9 cm,width=14cm}}
\caption{Atmospheric neutrino production mechanism.}
\label{fig:nuprod}
\end{center}
\end{figure}

\subsection{\bf $\mu$/e event rato for high energy neutrinos}
There is a qualitative explanation of the $\mu$ event suppression at high energy 
resulting from this study. The production of neutrinos in atmospheric showers 
is illustrated in Fig.\ref{fig:nuprod}. Charged pions produced through inelastic 
interactions of primary cosmic radiation decay as follows:
\begin{eqnarray}
& \pi^+\,\,\, \rightarrow & \mu^+ + \nu_\mu \nonumber \\
& & \hspace{0.2cm} \hookrightarrow e^+ + \nu_e + \overline{\nu_\mu} \nonumber \\
& \pi^-\,\,\, \rightarrow & \mu^- + \overline{\nu_\mu} \nonumber \\
& & \hspace{0.2cm} \hookrightarrow e^- + \overline{\nu_e} + \nu_\mu. \nonumber
\end{eqnarray}

\newpage 
Due to the Earth's magnetic field a charged particle is deflected before decay 
from its initial direction. 
The deflection, $\Delta \Theta$, is practically
independent of the particle momentum, p,

$$ \Delta \Theta ={L\over R} \sim {\gamma \over p}=
{{\sqrt{p^2+m^2}}\over{mp}}\simeq {const\over m},$$
where L is the decay path length and R is the radius of curvature of the
particle track in the magnetic field. 

For pions the mean decay length is short (c$\tau_\pi$=7.8 m) and, consequently,
the mean deflection is very small ($\Delta \Theta_{\pi} \simeq $ 0.4 mrad).
For muons the mean decay length is about 100 times longer (c$\tau_\mu$=658.6 m) 
resulting in a mean muon deflection before decay, $\Delta \Theta_{\mu} \simeq 3^o$.
Albeit relatively small, this deflection is sufficient to direct most of neutrinos
from atmospheric muon decays towards the ground, even allowing for the Earth's
spherical shape.   

Thus, the neutrinos from $\pi^+\rightarrow \mu^+ \nu_{\mu}$ and 
$\pi^-\rightarrow \mu^- \bar{\nu}_{\mu}$ decays of pions produced in the atmosphere 
and going nearly parallel to horizon never reach the surface of the Earth.
On the other hand (a half of) neutrinos from muon decays are emitted towards 
the Earth. Assuming equal flux from the East and the West, which is the case
for high energy neutrinos produced by high energy cosmic rays, one gets for the 
nearly horizontal flux: $\nu_{\mu}$ and $\bar{\nu_e}$ from the West and
$\bar{\nu_{\mu}}$ and $\nu_e$ from the East (see Fig.\ref{fig:nuprod}), and for 
the muon to electron event ratio:
$${\mu \over e}={{\nu_{\mu}+\bar{\nu_{\mu}}} \over {\bar{\nu_e}+\nu_e}}=1.$$

All neutrinos from the vertical direction reach the Earth's surface. This gives for 
the event ratio:
$${\mu \over e}= {{(\nu_{\mu}+\bar{\nu_{\mu}})+(\bar{\nu_{\mu}}+\nu_{\mu})}\over
{\bar{\nu_e}+\nu_e}}=2$$
in agreement with results of the present calculations. 
    
\subsection{\bf Hadronic interaction models}
Different groups use different models to simulate production of secondaries
in the atmospheric showers. Independent of the method used, all models are 
eventually fine tuned using as much available experimental data as possible.
The GHEISHA code of simulation of hadronic interactions used in this study
is not an exception.

There is a certain amount of criticism of GHEISHA in connection with neutrino flux 
calculations \cite{bib-BattF,bib-Wentz1,bib-Wentz} based on direct model tests presented in 
\cite{bib-Carmi,bib-Ferra}. 

The inability of GHEISHA to predict the properties of residual nuclear fragments and the related 
problems with momentum/energy conservation may be of importance in calorimetry, but have no 
pertinence to the secondaries responsible for production of charged leptons and neutrinos in the
energy range considered in this study.   

The results of \cite{bib-Carmi} are relevant for proton interactions with relatively heavy
(A$>$27) nuclei at proton energies $\rm E_{lab}<$1 GeV i.e. the region of energies 
too low to affect atmospheric neutrino fluxes.

In \cite{bib-Ferra} the interactions of pions with several nuclei are studied. 
It is acknowledged that as far as secondary pion multiplicity and Feynman x distributions are 
concerned GHEISHA behaves well up to the pion energies as high as 100 GeV. At the same time there 
are some deviations from experimental data, especially for heavy nuclei, in transverse momenta
and rapidity distributions of the secondaries. These deviations can not 
seriously influence the results of atmospheric shower simulations.           

An atmospheric shower is developing in the medium of rapidly changing density. A particle
arriving vertically with respect to the Earth's surface pass through an equivalent
of one nuclear interaction length when it reaches an altitude of about 17 km. The equivalent
of second interaction length is attained at about 13 km. At these altitudes the density of
atmosphere is 1/10 and 1/5 of air density at the ground level respectively. This corresponds
to nuclear interaction length of about 13 km and 3.5 km respectively.
A pion of energy as high as 20 GeV has a mean decay length of only about 1 km. 
Consequently, charged pions produced in atmospheric showers mostly decay in flight and do not
interact with atmosphere. For the neutrino fluxes, it means that the effect
of inaccuracy in angular distributions of secondary pions reported in \cite{bib-Ferra} is 
washed out by the change in direction of pions in the Earth's magnetic field and the 
subsequent pion decay.

\section{Conclusions}
Our calculations are based on the up-to-date data on the cosmic ray fluxes measured 
by the AMS detector \cite{bib-AMS}. The atmospheric showers are simulated with
as little simplifications as possible at all stages of primary and secondary particle 
travel and interaction with the atmosphere.
The difference in angular distributions of the ratios of fluxes of different kinds of 
neutrino at different sites and their dependence on the neutrino energy reported in 
\cite{bib-PLY} is confirmed on the basis of higher statistics.
The angular distributions and the neutrino event rates obtained in this study suggest a
possibility to describe the experimental results on the high energy atmospheric neutrino 
fluxes \cite{bib-Skam1,bib-MACRO} without evoking 
muon neutrino oscillations.

The good agreement between charged secondary particle spectra produced in this 
calculation and experimental data from AMS \cite{bib-AMS} and CAPRICE \cite{bib-CAPRICE}        
supports the viability of the result on atmospheric neutrino fluxes. 

{\bf Acknowlegements.} I'm grateful to Prof.Yu.Galaktionov for very useful and stimulating 
discussions.

\newpage

\newpage    
%\tableofcontents

   % gets the list of bibliography files

\end{document}